%% file: final2.tex
\documentclass{aa}
\usepackage{longtable,lscape}
\usepackage{graphicx}
\usepackage{natbib}
\bibpunct{(}{)}{;}{a}{}{,}
\usepackage{txfonts}
\usepackage{color}

\begin{document}
\def\cd{d$^{-1}$}
\def\cds{d$^{-1}$\,}
\def\rr{CoRoT 101128793}
\def\fB{$f_m$}
\def\fu{$f_0$}
\def\fd{$f_1$}
\def\ft{$f_2$}

\title{CoRoT light curves of RR Lyrae stars
\thanks{
The CoRoT space mission was developed and  is operated by the French
space agency CNES, with participation of ESA's RSSD and Science Programmes,
Austria, Belgium, Brazil, Germany, and Spain.}$^,$
{\thanks{The CoRoT timeseries, Tables 1 and 2 are available in electronic form
at the CDS via anonymous ftp to cdsarc.u-strasbg.fr (130.79.128.5)
or via http://cdsweb.u-strasbg.fr/cgi-bin/qcat?J/A+A/vol/page}
%\thanks{\ep{Tables 1 and 2 are available in the electronic edition of the journal at \tt{http://www.aanda.org}}}
}}
\subtitle{\rr: long--term changes in the Blazhko effect and excitation of additional modes 
}
%\author{E.~Poretti\inst{1}\fnmsep\thanks{\email{ennio.poretti@brera.inaf.it}}
\author{E.~Poretti\inst{1}
        \and
             M.~Papar\'o\inst{2}
        \and
             M.~Deleuil\inst{3}
        \and
             M.~Chadid\inst{4}
        \and
             K.~Kolenberg\inst{5}
        \and
             R.~Szab\'o\inst{2}
        \and
             J.M.~Benk\H{o}\inst{2}
        \and
             E.~Chapellier\inst{4}
\and
             E.~Guggenberger\inst{5}
\and 
         J.F.~Le~Borgne\inst{6}
\and
     F.~Rostagni\inst{4}
\and
             H.~ Trinquet\inst{4}
\and
       M.~Auvergne\inst{7}
\and
       A.~Baglin\inst{7}
\and
      L.M.~Sarro\inst{8}
\and
       W.W.~Weiss \inst{5}
}
% \offprints{E. Poretti}

 \institute{
   INAF -- Osservatorio Astronomico di Brera, Via E. Bianchi 46, 23807 Merate (LC), Italy\\
\email{ennio.poretti@brera.inaf.it}
\and
   Konkoly Observatory of the Hungarian Academy of Sciences, PO Box 67, H-1525 Budapest, Hungary
\and
   LAM, UMR 6110, CNRS/Univ. de Provence, 38 rue F. Joliot-Curie, 13388 Marseille, France
\and
   Observatoire de la C\^ote d'Azur, Universit\'e Nice Sophia-Antipolis, UMR 6525, Parc
   Valrose, 06108 Nice Cedex 02, France
\and
   Institute of Astronomy, University of Vienna, T\"urkenschanzstrasse 17, A-1180 Vienna, Austria
\and
Laboratoire d'Astrophysique de Toulouse-Tarbes, Universit\'e de Toulouse,
CNRS, 14 Av. Edouard Belin, 31400 Toulouse, France
\and
    LESIA, Universit\'e Pierre et Marie Curie, Universit\'e Denis Diderot, Observatoire de
    Paris, 92195 Meudon Cedex, France
\and
Dpt. de Inteligencia Artificial, UNED, Juan del Rosal 16, 28040 Madrid, Spain 
}
  \date{Received, accepted}
  \abstract
  {
The CoRoT (Convection, Rotation and planetary Transits) space mission provides 
a valuable opportunity to monitor stars with uninterrupted time sampling for up to 150 days at a time.
The study of RR Lyrae stars, performed in the framework of the Additional
Programmes belonging to the exoplanetary field, will particularly benefit from
such dense, long-duration monitoring.
}
{
The Blazhko effect in RR Lyrae stars is a long-standing, unsolved problem of stellar
astrophysics. We used the CoRoT data of the new RR Lyrae variable \rr\, 
(\fu=2.119~\cd, $P$=0.4719296~d) to provide us with 
more detailed observational facts to understand the physical process behind the phenomenon.
}
{The CoRoT data were corrected for one jump and the long-term drift.
We applied different period-finding techniques to the corrected timeseries to 
investigate amplitude and phase modulation. We detected 79 frequencies in the light
curve of \rr. They have been  identified as the main frequency \fu\, and its harmonics, 
two independent terms, the terms related to
the Blazhko frequency \fB, and several combination terms.} 
  {A Blazhko frequency \fB=0.056~\cds  and a triplet structure 
around the fundamental radial mode and harmonics were detected, as well as  
a long-term variability of the Blazhko modulation.
Indeed, the amplitude of the main oscillation is decreasing along the
CoRoT survey. The Blazhko modulation is one of the smallest observed in
RR Lyrae stars.
Moreover, the additional modes \fd=3.630~ and \ft=3.159~\cds are detected. Taking 
its ratio with the fundamental radial mode into account, the term \fd\, could be the identified
as the second radial overtone.
Detecting of these modes
in horizontal branch stars is  a new result obtained by CoRoT. 
}
{}

\titlerunning{CoRoT light curves of RR Lyrae stars}
\authorrunning{E. Poretti et al.}

   \maketitle

\section{Introduction}

The pulsation of RR~Lyrae stars is  paramount for
advancing in several fields of stellar physics. 
\citet{marcella} emphasizes  how we can reproduce all the relevant
observables of the radial pulsation including only
non-local, time-dependent treatment of the convection in nonlinear models. In particular,
pulsational models are able to reproduce the correlation between
the periods and the absolute magnitudes in the near infrared bands
\citep{bono}. The model-fitting technique \citep{marcella} applied
to a sample of  RR Lyrae stars in the Large Magellanic Cloud  
was very useful to fix the problem of  the distance scale \citep{lmc}. 
Because they have been observed since the end of the XIXth century, RR Lyrae stars are also promising targets
for studying stellar evolution in real time \citep{flb}.

What has not yet been understood in RR Lyrae stars is the Blazhko effect, a periodic modulation
of both the amplitudes and phases of the main pulsational mode. 
Different mechanisms have been
proposed to explain the phenomenon: the resonance model between nonradial modes of low degree
and the main radial mode \citep{dzie}, the oblique pulsator  model in which the rotational
axis does not coincide with the magnetic axis  \citep{don, shiba}, 
and the action of a turbulent convective dynamo in the lower envelope of the star \citep{sto}.  
\citet{geza} 
reviews these models and points out why we cannot definitely accept any of these
explanations. It seems well--established that Blazhko RR Lyrae stars do not
show any strong magnetic field \citep{mer,kk}.
The observation of Blazhko RR Lyrae stars was performed with remarkable success by
means of extensive ground--based surveys. Well--defined findings
(e.g., changes in the Blazhko period, modulation features, systematic changes in the
global mean physical parameters, high--order multiplets, long--term changes) have recently been
obtained on RR Lyr itself \citep{rrlyr}, MW Lyr \citep{mw}, XZ Cyg \citep{xz}, RR Gem \citep{gem,gem2},
and DM Cyg \citep{dmcyg}.

The Additional Programmes  in the exoplanetary science case of the CoRoT 
(COnvection, ROtation and planetary Transits; \citealt{esa3}) space mission
were focused on specific classes of stars with the aim of supplying a new and powerful
tool for deciphering the physical reasons for their variability \citep{esa4}.
RR Lyrae stars are being studied in the framework of the international RR~Lyrae-CoRoTeam\footnote
{The dedicated
website is http://fizeau.unice.fr/corot.}. Preliminary results 
were presented by \citet{chasf}, and the 
potential of the 150--day  continuous monitoring of an RR Lyrae star has been demonstrated 
by the case of V1127~Aql \citep{aql}, not previously known as   
a Blazhko variable. Very high--order modulation sidepeaks were detected, up to
the sepdecaplet structure. Additional modes have also been detected and interpreted as nonradial
modes or secondary modulation. As the Blazhko effect remains misunderstood
in its physical nature, we can look at the CoRoT data as a new opportunity for
providing the observational facts we need to shed new light on it.

\begin{figure}[]
\begin{center}
\includegraphics[width=\columnwidth,height=\columnwidth]{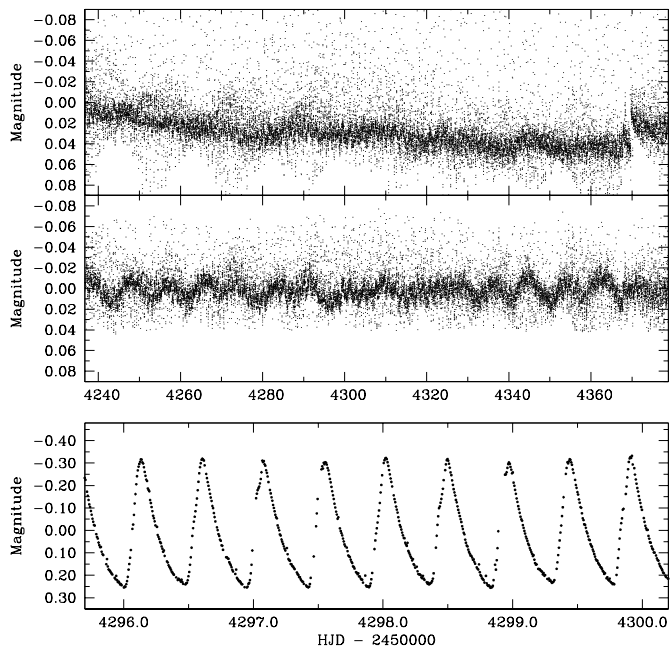}
%\vskip 8.5truecm
\caption{\footnotesize {\it Top panel:} the light curve obtained by removing the main frequency and its
harmonics from the original data showing a long-term drift and a jump. {\it Middle panel:} the light curve
of the residuals is corrected from the long-term drift and the jump.
{\it Bottom panel:} the final light curve of \rr\, is an example of the continuous, excellent quality monitoring
of stars in the CoRoT exoplanetary field.}
\label{rid}
\end{center}
\end{figure}

\section{The CoRoT data}
\rr$\equiv$USNOA2~0900-15089357 
($\alpha$ = $19^{\mathrm h}$\,$26^{\mathrm m}$\,$37\fs{33}$,
$\delta$ = +01\degr\,13\arcmin\,35\farcs{05}, J2000)
is a 16$^{\rm th}$--mag star ($V$=15.93, $B-V$=+0.89, \citealt{exodat}) 
in the constellation of Aquila.
Its  variability was discovered during the first Long Run in the
centre direction (LRc01), carried out continuously from  May 15 to  
October 14, 2007, i.e., for 142~d.
There is no relevant contamination from nearby stars, since the brightest  
star included in the CoRoT mask is 3.0~mag fainter than \rr\, in $V$ light
\citep{exodat}.  
The exposure time in the CoRoT exoplanetary
channel was 512~sec and this time remained constant all over the LRc01.
Thanks to its very high duty cycle, CoRoT collected 23922 data points, 
and the spectral window is free from any relevant alias structure. 
The star was classified as an RR Lyrae variable  by the 
``CoRoT Variability Classifier" automated supervised method \citep{cvc}
and then confirmed by human inspection of the light curve. 
\rr, located close to the direction of the galactic centre,
is therefore heavily reddened.

The absolute CoRoT photometry is affected by jumps, outliers,
and a long-term drift. It is very hard to detect jumps in the original data of an 
RR Lyrae variable, since they have a small amplitude (few 0.01-mag) and 
are not discernible in a light curve having an amplitude of several tenths of a magnitude.
As a matter of fact, we could detect a jump of 0.032~mag at JD~2454369.7
only a posteriori, after having performed the preliminary frequency analysis
of the original data. Indeed, only
the residuals obtained by subtracting the main frequency \fu\, and its harmonics 
from the original CoRoT data clearly show the jump.
We re-aligned the whole dataset  after removing the few corrupted measurements on the jump
(Fig.~\ref{rid}, top and middle panels).

In addition to the jump, some oscillations and a continuous drift are clearly visible in the 
top panel of Fig.~\ref{rid}. The oscillations have a stellar origin (see Sect.~\ref{blaz}), but
the drift is an instrumental effect \citep{flight}, so 
it should be removed before performing the frequency analysis.
Different detrending algorithms can
be used, based on  moving means or polynomial fits. After several trials, we 
removed the drift by calculating the mean magnitudes of the least--squares fits
of four consecutive cycles (i.e., 1.88~d). The main frequency and its harmonics were used, as
in the previous step. At that point, the value of the mean magnitude was interpolated at the
time of each observation and then subtracted from the original data.
During this  analysis we also removed  the most obvious outliers.
The final CoRoT timeseries is available in electronic form at the CDS. 
The re-aligned, de-jumped light curve  disclosed  the multiperiodic behaviour of \rr:  continuous
oscillations  are clearly visible in the light curve prewhitened with \fu\, and harmonics
(Fig.~\ref{rid}, middle panel) and in the light curve of the original data
(a portion is shown in Fig.~\ref{rid}, bottom panel). 

The subsequent frequency analysis was performed by using different packages such as
Period04 \citep{Lenz05},  MuFrAn \citep{Kollath90},
and the iterative sine--wave fitting \citep{vani}.
The different algorithms  led to the same results with only marginal  
differences at higher orders. We present here the results of the iterative--sine
wave fitting, with a complementary frequency refinement obtained by means of   
the MTRAP algorithm \citep{mtrap}.

The realigned dataset was first analysed to 
search for the effects of the orbital frequency. 
Several frequencies were found at the orbital frequency $f_{orb}$=13.97~d$^{-1}$
and harmonics. Moreover, the term $f_{sid}$=1.0027~d$^{-1}$ was found. This
perturbation comes from the passage of  the satellite over the South Atlantic 
Anomaly (SAA). Since it occurs twice a day, the harmonic 2$f_{sid}$ is much stronger
than $f_{sid}$, which corresponds to the passage of the satellite over the SAA 
on the same side of the
Earth with respect to the Sun. The effects of these passages on the onboard
instrumentation are described by \citet{flight}. They originate frequencies  at   
\begin{equation}
f_{o,s}=k_1~f_{orb} \pm k_2 f_{sid}
\end{equation}
with $0\le k_1 \le 7$ and $-4\le k_2 \le 4$. The strongest terms are
27.94 and 41.91~d$^{-1}$, i.e., $(k_1,k_2)=(2,0)$ and (3,0), respectively. The usually adopted 
technique of prewhitening the input data with the frequencies $f_{o,s}$ did not
correct for the instrumental effects in a satisfactory way. The  orientation of the 
CoRoT orbital plane with respect to the Earth--Sun line continuously changed over the course of
the long run. Therefore, the environmental conditions  
(e.g., the eclipse effects on the electronics units,
the eclipse durations, the difference in the Earth's albedo of the overflown regions;
see Sect.~3 in \citealt{flight}) are affecting the CoRoT photometry in a complicated 
way.

\begin{figure}[]
\begin{center}
\includegraphics[width=\columnwidth,height=\columnwidth]{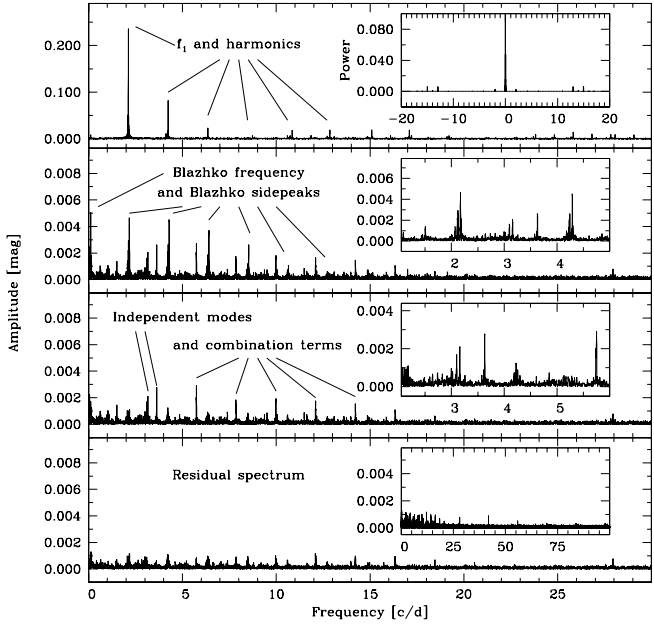}
%\vskip 8.5truecm
\caption{\footnotesize Subsequent steps in the detection of frequencies in the amplitude spectra of  \rr.}
\label{spettro}
\end{center}
\end{figure}

\section{The frequency content}
By using the packages previously mentioned,
we identified 79 components of stellar origin,
in addition to the $f_{o,s}$ frequencies and to the spurious peaks at very low frequencies, 
i.e., residuals of the long--term drift of the sensitivity drift of the CCDs.
They  can be divided into four categories:
\begin{enumerate}
\item the main frequency \fu\, and its harmonics; 
\item the terms related to the Blazhko frequency \fB;
\item other independent terms;
\item the combination terms.
\end{enumerate}
Figure~\ref{spettro} describes the different steps in the frequency detection.
The spectrum in the top panel brings out the main frequency
\fu=2.119~\cds and its harmonics. The spectral window (inserted box) is almost
free of aliases, and the peaks located at $f_{orb}$ and 2$f_{sid}$ 
are too low to produce any significant effect. When \fu\, and
harmonics are removed, the couple of  sidepeaks (\fu$\pm$\fB, with \fB=0.056~\cd) 
due to the Blazhko effect 
becomes the most prominent structure (second panel, the zoom 
around \fu\, and 2\fu\, is shown in the inserted box). 

The most intriguing peaks stand out 
in the region $3~-~4$~\cds  after subtracting  the Blazhko sidepeaks 
(third panel and inserted box). The highest peaks in the third panel of Fig.~\ref{spettro}
are at \fd=3.157~\cds and \ft=3.630~\cd. 
They show linear combination with  \fu\, and harmonics and are therefore
intrinsic to the RR Lyrae star. They provide evidence of excited modes 
other than the fundamental radial mode \fu.

The residual  spectrum does not show any other structure, except
an excess of signal still centred on the largest amplitude modes
and on the orbital frequencies of the satellite
(Fig.~\ref{spettro}, bottom panel).
After removing the 79 frequencies, 
the average noise level resulted in 
$7\times~10^{-5}$~mag in the 0-100~\cds region of the residual spectrum
(inserted box in Fig.~\ref{spettro}, bottom panel). 
The lowest detected amplitude among the 79 frequencies led to 0.36~mmag,
i.e., 5 times the level of the overall final noise.
We note that at each step of the process in  
frequency detection we calculated the local noise centred on the detected peak,
and we always got SNR$>$3.5. 
This threshold was retained to accept  a combination term,
while independent terms have much higher SNR (17.45 and 9.25 for \fd\, and \ft, 
respectively). 

The final solution of the CoRoT light curve  was calculated by means of a 
cosine series (T$_0$=2454308.2168) and their least--squares parameters, together with the local
SNRs, are listed in Table~\ref{solution}. The listed values of the  frequencies are
corresponding to the highest peaks in the amplitude spectrum. The  values calculated from
the four independent frequencies and the identification listed in the last column of
Table~\ref{solution} (the so-called locked solution, obtained by using the MTRAP
algorithm, \citealt{mtrap})
are generally in excellent agreement
(\fu=2.118977, \fd=3.630499, \fd=3.156776, and \fB=0.00550~\cd).
The observed  discrepancies
are probably due to the non--equidistance of the triplet structures
and to other terms hidden in the residual noise.
As an example, a third independent frequency is probably present close to 3.00~\cd, but we
cannot identify it unambiguously. If this $f_3$ term were real, then some combination
terms should be changed by substituting, e.g.,  2\fd\ with 2\fu$+f_4$. 
The solution with all independent frequencies  gives the same residual rms of the
solution with the locked frequencies (0.01006 and 0.01100 mag, respectively). These values
are mostly affected by the residual peaks described above.

\subsection{The main \fu\, term and its harmonics}
\begin{figure}[]
\begin{center}
%\vskip 8.5truecm
\includegraphics[width=\columnwidth,height=\columnwidth]{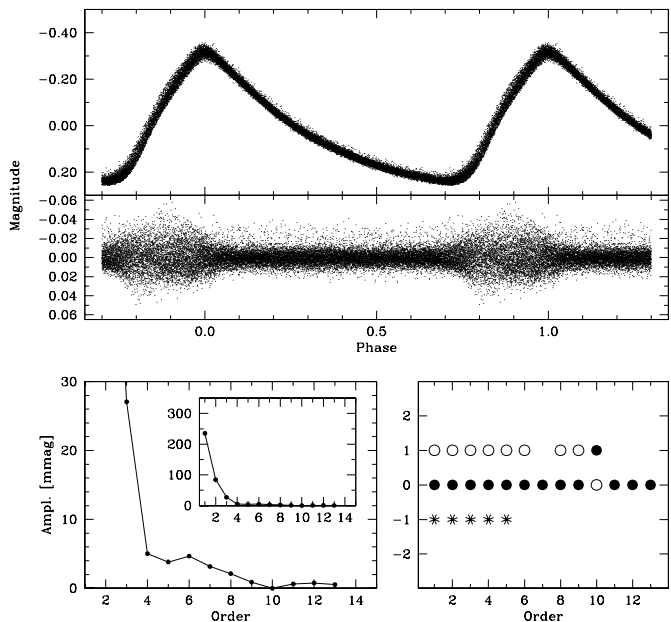}
\caption{\footnotesize
{\it Upper panel:} The CoRoT data folded with \fu, original (top) and after
subtracting the 79 frequencies (bottom). 
{\it Left panel, lower row:} The amplitudes of the $n$\,\fu\, terms.
{\it Right panel, lower row:} The observed triplet structure around  $n$\,\fu\, terms.
The filled circles indicate the component with the greatest amplitude in the triplet,
the empty circles the second in amplitude, the star the third.
}
\label{f1}
\end{center}
\end{figure}

The light curve on the \fu\, term is very asymmetric (Fig.~\ref{f1}, upper curve
in the top panel) and
harmonics up to 13\fu\,   are significant. Their amplitudes  are not 
monotonically decreasing:
the amplitude of 6\fu\,  is larger than for 5\fu, and that of 11\fu\, is larger than for 10\fu,
just before the final decline (Fig.~\ref{f1}, left panel in the bottom row).
Indeed, the light curve of \rr\, shows a couple of particularities, i.e., the bump near the
minimum often observed in RRab stars  and a change in slope on the rising branch. They are not
very pronounced, but still discernible in the light curve (Fig.~\ref{rid}, bottom panel). 
The fit of these small particularities  requires  a more relevant contribution 
from the highest harmonics than in the case of
smooth light curves. Moreover, the change in slope 
does not repeat in  a regular way, since the plot of the residuals
(Fig.~\ref{f1}, lower curve in the top panel) shows a wide spread in this phase interval. The non--white
distribution of the photometric residuals is the cause of the small bunches of frequencies  observed
in the residual spectrum. The Blazhko variables RR~Gem and DM~Cyg show the same light curve shape and 
the same residual distribution as \rr\, \citep{gem,dmcyg}. 

The measurements around the maximum and minimum brightnesses were fitted by means of a least-squares 
polynomial. We obtained the ephemeris 

$
\begin{array}{lrl}
{\rm Max =  HJD}& 2454236.6752 & + 0.4719296\,\, {\rm x E}\\
               &    \pm0.0003 &\pm0.0000018  
\end{array}
$\\
when fitting the times of  maxima (Table~\ref{maxima}) by means of a least--squares line. 
The O-C values (differences between the observed and calculated times of maxima)
were determined by using this ephemeris.

\subsection{The Blazhko frequency \fB}\label{blaz}
As suggested by the residuals after subtracting the main oscillation (Fig.~\ref{rid}, middle panel),
there is a periodic change in the shape of the light curve, and
this change defines the Blazhko effect. 
The  Blazhko effect  translates
into symmetric sidepeaks of \fu\, and its harmonics in the frequency domain
(second panel in Fig.~\ref{spettro}).
In the case of  \rr, the sidepeaks are $n$~\fu$\pm$\fB\, triplets (Fig.~\ref{f1}, right panel in the bottom row). 

\begin{figure}[]
\begin{center}
%\vskip 8.5truecm
\includegraphics[width=\columnwidth,height=\columnwidth]{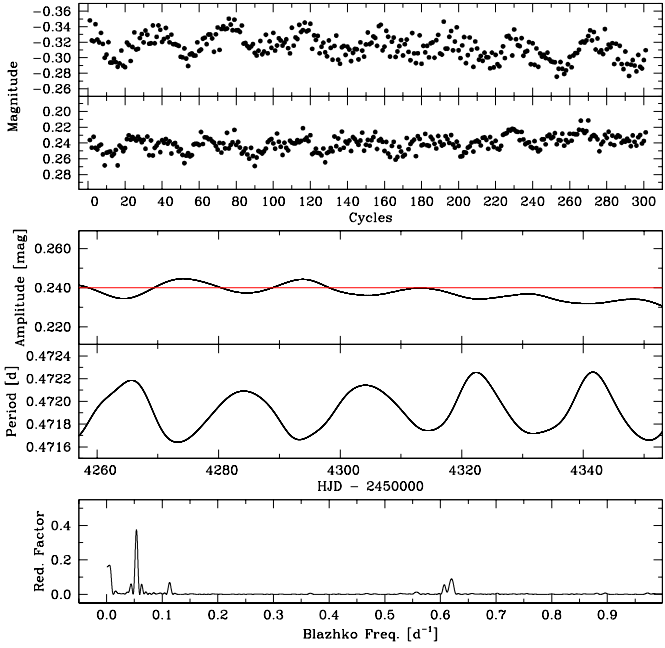}
\caption{\footnotesize
Evidence of the long--term variations in the \rr\, light curve.
{\it Top panel:} The magnitudes at maximum (upper curve) and minimum
(lower curve) brightness  in the different cycles. 
{\it Middle panel:} The amplitudes of the \fu\, component (upper curve) and 
the period values (lower curve). Reference line is displayed to show
the long term variation in amplitude. 
{\it Bottom panel:} The power spectrum of the magnitudes at maximum brightness.
}
\label{magmax}
\end{center}
\end{figure}

We obtained an independent confirmation of the Blazhko frequency 
from the magnitudes at the maximum brightness (see above)  and from the application
of the analytic signal method \citep{analytic}.  
The magnitudes at maximum  oscillate in a peak-to-peak interval of 0.06~mag 
(Fig.~\ref{magmax}, top panel): the power spectrum  
unambiguously identifies  \fB=0.056~\cds (Fig.~\ref{magmax}, bottom panel). 
The instantaneous amplitudes and frequencies also vary with \fB\, (Fig.~\ref{magmax}, 
middle panel). The period variations, and consequently the O-C range, are  very small.
As a matter of fact, \rr\, shows the smallest period variation among the CoRoT RRab stars
(see Fig.~2 in \citealt{szabosf}).
Since the Blazhko effect is more evident in amplitude than in phase,
the cycle--to--cycle variations in the light curve are undetectable when folding
the data over \fu, also considering  the
perfect coverage in phase ensured by the CoRoT observations.
The Blazhko effect just causes a wider spread  of the points, while
the curve apparently remains very regular (Fig.~\ref{f1}, top panel).
The Blazhko effect of \rr\, seems to be  particular since the harmonic 2\fB\, has an
amplitude greater than \fB\, (Table~\ref{solution} and middle panel of
Fig.~\ref{rid}).  This particularity could reflect  the different forms 
in  which the Blazhko effect can occur \citep[see Table~1 in][]{szabosf}.
However, 
we should also consider that the frequency and amplitude values of \fB\, and 2\fB\,
could be affected
by the slightly different separations between the sidepeaks of the triplets
(Table~\ref{solution}) and by the correction of the low--frequency drift.

We observe a peak  close to zero (Fig.~\ref{magmax}, bottom panel) in 
the power spectrum of the magnitudes at maximum. This   
suggests that there is a very long--term variation,  
on a timescale longer than covered by the CoRoT data.
The  complicated behaviour of the light curve  is made clear by the comparison
between maxima and minima  (Fig.~\ref{magmax}, upper  panel). The
range in the magnitudes at minimum is about half that at  the maximum. 
Also the amplitudes of the \fu\, component (calculated both as instantaneous values and 
by subdividing the timeseries in pulsational cycles)
show the decreasing  trend underlying the Blazhko periodicity 
(Fig.~\ref{magmax}, middle panel).

The long--term change is a further complication of  the frequency analysis 
\citep{Benko09, Szeidl09}.
Together with the change in slope on the rising branch, it 
causes the peaks  around \fu\, and  its harmonics   
in the residual spectrum (Fig.~\ref{spettro}, bottom panel). 

\subsection{The independent terms}
\subsubsection{\fd=3.630~\cd}
The first peak not related to \fu\, and \fB\, is found
at \fd=3.630~\cd. The light curve related to this  periodicity is slightly
asymmetrical, since we found  a small-amplitude first harmonic 2\fd. 
It also shows several combination terms with $n$~\fu\, and $n$\,\fu$\pm$\fB.
However, \fd\, is not affected by the Blazhko effect, since we did not detect terms of
the form \fd$\pm$\fB. 

\begin{figure}[]
\begin{center}
%\vskip 8.5truecm
\includegraphics[width=\columnwidth,height=\columnwidth,angle=270]{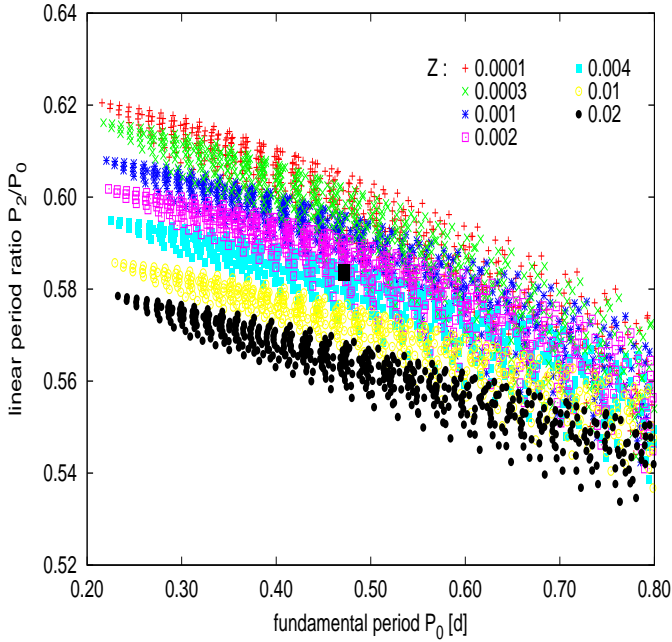}
\caption{Petersen diagram based on linear convective RR Lyrae models. The
symbols denote different metallicities. The black square  shows the position
of \rr, assuming that the frequency $f_2$ is the second overtone radial mode.}
\label{model}
\end{center}
\end{figure}

The ratio \fu/\fd=0.584 is very close to what is expected between
the fundamental radial mode and the second overtone. To verify this possibility
from a theoretical point of view, we computed linear RR Lyrae model grids on an  
extremely large parameter space ($L$=40, 50, 60 and
70~L$_\odot$, $M$=0.50--0.80~M$_\odot$ with $\Delta$M=0.05~M$_\odot$,
$T_{\rm eff}$=5000--8000~K,
$\Delta T_{\rm eff}$=100~K, $Z=$0.001, 0.003, 0.01, 0.02 and 0.04). 
The other adopted parameters were standard RR~Lyrae parameters (see \citealt{Szabo94}).
Nonlinearity introduces a negligible difference in the periods and period ratios.
The Petersen diagram for different metallicities is shown in Fig.~\ref{model}.
The period ratio is fully compatible with an identification of \fd\, as the second 
radial overtone. In such a case, the models suggest a $Z$-metallicity of 0.002-0.004.
Assuming a ratio of 0.74 between fundamental and first overtone radial modes, the latter
should be around 2.863~\cd, but the frequency  spectrum
does not show any significant peak at this value. 

\subsubsection{\ft=3.159 \cd}
The amplitude of \ft=3.159 \cd\, is only a bit smaller than that of \fd\, (0.0021 and 0.0028 mag, respectively)
and at the same level of that of the 8\fu\, harmonic. 
The ratio \ft/\fu=1.4908$\sim$3/2 could be the signature of the period doubling bifurcation \citep{pawel}
first noticed in some RR Lyrae stars observed with  {\it Kepler} \citep{kkapj}.
%Such a phenomenon could involve the resonance between the fundamental radial mode and another 
%(radial \ep{or}  non-radial) mode.  

Another characteristic of \ft\, is to be flanked by a Blazhko frequency at \ft$-$\fB. 
This occurrence can have different explanations : {\it i)} the Blazhko variability also
affects \ft; 
{\it ii)}  it is a coincidence, and \ft\, and \ft$-$\fB\, are actually two independent modes;
{\it iii)} \ft$-$\fB\, is a mere combination term, such as the difference between \fu$+$\ft\, 
and \fu$+$\fB.
We can try to disentangle the matter by examining the three possibilities.
If \ft\, shows  the Blazhko effect, it is strange that we do not detect \ft$+$\fB\, since we
expect the sum term to have an amplitude greater than the difference term (see
Fig.~\ref{f1}, right panel in the bottom row). The coincidence is also improbable, since the frequency
spectrum is not very rich.  The resonance mechanism is possible, but that it involves the Blazhko
term makes it a very particular
case. Finally, combination terms with $n$~\fu, $n$\,\fu$\pm$\fB,
and \fd\, are detected. In particular, we found unusual combinations, such as \fu+\fd+\ft\, 
and \fu+\fB+\fd+\ft,  and they display a good SNR (5.4 and 4.6, respectively).
Therefore, the hypothesis of a particular combination term seems the most plausible.

\section{Discussion}
The continuous, long monitoring offered by space photometry  
is a new observational tool to understand the pulsational behaviour of RR~Lyrae stars. 
Through such data, cycle--to--cycle variations can be clearly
pointed out. Indeed, the Blazhko modulation of \rr\, is one of the smallest ever observed
in RR Lyrae stars
\citep{konko}.

\subsection{The Blazhko effect}
CoRoT data already demonstrated  that the Blazhko cycle of V1127 Aql
is  changing on a timescale of 143~d: the shift is much more evident in time than in magnitude 
(see Fig.~14 in \citealt{aql}). The Blazhko effect of \rr\, is much smaller than that
of V1127 Aql (0.06 mag vs. 0.35 mag in the full range of magnitudes at maximum, 0.02~p vs. 0.17~p
in the phases of maximum). 

Notwithstanding this small effect, the trend observed in the magnitudes at maximum and at
minimum (Fig.~\ref{magmax}, upper  panel) supports a long--term change.  
The best observational evidence for a long--term change in the Blazhko period is given by RR Lyr itself.
Ground--based photometry collected on several decades shows a decrease from 40.8~d to 38.8~d \citep{rrlyr}. 
The modulation amplitude of RR Gem was also subjected to strong variations   
from the undetectable level (less than 0.04~mag in maximum brightness) to about 0.20 mag
on  a time baseline of 70~years \citep{gem2}.
In the case of the Blazhko effect of MW Lyrae (\fB=0.060~\cd), \citet{mw} 
put in evidence secondary peaks  around the main pulsation terms separated by a periodicity
comparable with the time baseline, tentatively $\sim$500~d.
Therefore, it seems that long--term changes are occurring in  
Blazhko RR Lyrae variables, and they can be detected when monitored in an intensive and/or continuous way.

\subsection{The excitation of additional modes}
The case of \rr\, supplies new evidence of the excitation of additional modes in RR Lyrae stars.
The two frequencies \fd\, and \ft\, are not related with the Blazhko or another modulation, as
it could be for V1127 Aql \citep{aql}. The frequency \fd\, could be typified as the second overtone radial mode,
while the nature of \ft\, is still unclear.
We immediately note that also in the case of V1127~Aql we found frequency
ratios compatible with that between fundamental and second overtone radial modes, i.e.,
2.8090/4.8254=0.582,  and with the possible period doubling bifurcation, i.e., 4.1916/2.8090=1.492.
Moreover, the frequency values \fd\, and \ft\, are in the same
interval of the nine additional modes detected in the frequency spectrum of V1127 Aql (3.64--4.82 \cd).

It is interesting to revisit the results obtained by \citet{mw} on MW Lyr. Those authors
identified four frequencies in the 3.27--6.78~\cds interval (3.2701, 4.2738, 5.7847, and 
6.7885~\cd)\footnote {We note that there is a 1~\cds spacing
between the terms of the first couple and between those of the second one. 
This spacing is always a bit suspicious for observations collected mostly from one site,
but we consider the frequency detection performed by \citet{mw} as a well-established one.}    
as  combination terms having the form $n\,f_0\,\pm\,12.5\,f_m$ (where $f_0$=2.5146~\cds   is the
main pulsation mode and $f_m$=0.0604~\cds the Blazhko frequency). 
Since the 12.5~$f_m$ spacing remains unexplained,
we propose an alternative solution based on the  additional modes 
$f_1$=3.2701~\cds and $f_2$=4.2738~\cds and the combination terms $f_0+f_1$=5.7847~\cd
and $f_0+f_2$=6.7884~\cd. We find for the third time a frequency ratio 
($f_0/f_2$=0.588) that could be explained with the ratio between the fundamental and the second overtone radial modes. 

These mode identifications  
are a new contribution to the debate on the excitation of the second overtone in RR Lyr stars
\citep[e.g.,][]{alcock, walker, kaluzny, soszinski, olech}.
We also note that
the excitation of non-consecutive radial modes would be a new result for RR Lyr
stars, so far sporadically observed in Cepheids \citep{triple}.

\section{Conclusions}
The second detailed analysis of the CoRoT data on RR Lyrae variables allowed us to advance in
the definition of their pulsational characteristics. It is confirmed  
that the Blazhko effect can span different ranges in the variations, both absolute and relative,
of the amplitude and phase modulations. Moreover,  there
is new evidence that the Blazhko period is subjected to long--term variations, as can 
be directly detected  from the consecutive cycles observed in the CoRoT LRc01.
The  mechanisms invoked to explain the Blazhko
effect should  reproduce ``the strictly regular behavior of the modulation observed in 
many Blazhko stars" \citep
{konkoly}. This requirement should now be reconsidered in a slightly different way. The real mechanism
must be able to reproduce both the regular
structure of the sidepeaks in the frequency spectra and 
the observed  variability on a long--term scale. 
In this context, it can be useful to stress that 
\rr, similar to DM Cyg \citep{dmcyg} and RR~Gem \citep{gem}, shows a bump on the rising branch of the
light curve.  These bumps are probably connected with hypersonic shock waves \citep{shock},
and the spreading of the residuals suggests a 
link between pulsation and atmosphere dynamics. More precisely, this
could be the clue to an interaction between the Blazhko phenomenon and the atmosphere's dynamics
\citep{bump}, since
monoperiodic RR Lyrae stars also have very regular light curves  in the presence of this bump
\citep{ogle}.  

The other relevant result disclosed by the CoRoT data is the excitation of additional 
modes. A reanalysis of the V1127 Aql and MW~Lyr cases
seems to indicate  that there is a narrow frequency interval where a few modes are excited. 
The recurrence
of the ratio 0.58--0.59 between one of these modes and the fundamental radial mode suggests
the possibility of the (preferred) excitation of the second overtone.
The possible interplay between this type of double--mode pulsation and the Blazhko effect
deserves further theoretical investigation. 
The Blazhko effect does not modulate the \fd\, term and this is particularly relevant
in this scenario.
%The non--modulation of the \fd\, term is of particular relevance in this scenario.
Moreover, CoRoT detected other very significant 
peaks in the oscillation spectra of V1127 Aql and \rr, thus disclosing the evidence that nonradial
modes are excited in horizontal branch stars. The theoretical prediction of these modes is the
new challenge to the pulsation models of RR Lyrae stars launched by  CoRoT.
\begin{acknowledgements}
This research has made use of the Exo-Dat database, operated at LAM-OAMP, Marseille, France, 
on behalf of the CoRoT/Exoplanet programme.
MC thanks F.~Baudin and J.~Debosscher for their help on the data reduction.
JMB, MP, and RSz acknowledge the support of the ESA PECS projects No.~98022 \& 98114.
KK and EG acknowledge the projects FWF~T359 and FWF~P19962, and EP the Italian ESS
project (contract ASI/INAF/I/015/07/0, WP~03170) for financial support.
\end{acknowledgements}

\vfill
\eject
\clearpage
%\end{document}
%\begin{appendix}
%\end{appendix}

\Online

\longtab{1}{
%\begin{longtable}{lllll}
\begin{longtable}{rrrrc}
\caption{Fourier amplitudes, phases, signal-to-noise ratio, and
identification of the frequencies detected in the 
CoRoT data of the star \rr.}\label{solution}
\\
\hline\hline
\noalign{\smallskip}
\multicolumn{1}{c}{Frequency} &
\multicolumn{1}{c}{Amplitude} &
\multicolumn{1}{c}{Phase}  &
\multicolumn{1}{c}{SNR}  &
\multicolumn{1}{c}{ID}  \\
\multicolumn{1}{c}{$\mathrm{[d^{-1}]}$} &
\multicolumn{1}{c}{[mag]} &
\multicolumn{1}{c}{[0,2$\pi$]}  \\
\noalign{\smallskip}
\hline
\endfirsthead
\caption{continued.}\\
\hline\hline
\noalign{\smallskip}
\multicolumn{1}{c}{Frequency} &
\multicolumn{1}{c}{Amplitude} &
\multicolumn{1}{c}{Phase}  &
\multicolumn{1}{c}{SNR}  &
\multicolumn{1}{c}{ID}  \\
\multicolumn{1}{c}{$\mathrm{[d^{-1}]}$} &
\multicolumn{1}{c}{[mag]} &
\multicolumn{1}{c}{[0,2$\pi$]}  \\
\noalign{\smallskip}
\hline
\noalign{\smallskip}
\endhead
\noalign{\smallskip}
\hline
\endfoot
\input{14941tb1.tex}
\end{longtable}
}

\longtab{2}{
\begin{longtable}{rccr c | rccr }
\caption{Times,  magnitudes, O--Cs of the maxima observed in the light curve of \rr. }\label{maxima}
\\
\hline\hline
\noalign{\smallskip}
\multicolumn{1}{c}{Cycle} & \multicolumn{1}{c}{Times}      & \multicolumn{1}{c}{Magnitude}& \multicolumn{1}{c}{O--C}  &&
\multicolumn{1}{c}{Cycle} & \multicolumn{1}{c}{Times}      & \multicolumn{1}{c}{Magnitude}& \multicolumn{1}{c}{O--C}  \\
\multicolumn{1}{c}{[E]}      & \multicolumn{1}{c}{of maximum} & \multicolumn{1}{c}{}         & \multicolumn{1}{c}{[d]} &&
\multicolumn{1}{c}{[E]}      & \multicolumn{1}{c}{of maximum} & \multicolumn{1}{c}{}         & \multicolumn{1}{c}{[d]} \\
\multicolumn{1}{c}{}      & \multicolumn{1}{c}{[HJD-2450000]} & \multicolumn{1}{c}{}      & \multicolumn{1}{c}{} &&
\multicolumn{1}{c}{}      & \multicolumn{1}{c}{[HJD-2450000]} & \multicolumn{1}{c}{}  \\
\noalign{\smallskip}
\hline
\noalign{\smallskip}
\endfirsthead
\caption{continued.}\\
\hline\hline
\noalign{\smallskip}
\multicolumn{1}{c}{Cycle} & \multicolumn{1}{c}{Times}      & \multicolumn{1}{c}{Magnitude}& \multicolumn{1}{c}{O--C}  &&
\multicolumn{1}{c}{Cycle} & \multicolumn{1}{c}{Times}      & \multicolumn{1}{c}{Magnitude}& \multicolumn{1}{c}{O--C}  \\
\multicolumn{1}{c}{[E]}      & \multicolumn{1}{c}{of maximum} & \multicolumn{1}{c}{}         & \multicolumn{1}{c}{[d]} &&
\multicolumn{1}{c}{[E]}      & \multicolumn{1}{c}{of maximum} & \multicolumn{1}{c}{}         & \multicolumn{1}{c}{[d]} \\
\multicolumn{1}{c}{}      & \multicolumn{1}{c}{[HJD-2450000]} & \multicolumn{1}{c}{}      & \multicolumn{1}{c}{} &&
\multicolumn{1}{c}{}      & \multicolumn{1}{c}{[HJD-2450000]} & \multicolumn{1}{c}{}  \\
\noalign{\smallskip}
\hline
\noalign{\smallskip}
\endhead
\noalign{\smallskip}
\hline
\endfoot
\input{14941tb2.tex}
\end{longtable}
}

\end{document}

%% file: 14941tb1.tex
\\
\multicolumn{5}{l}{Main frequency and harmonics} \\
\\
  2.1189511 &   0.235160 &   0.2021  & 83.170 &   $f_0$  \\
  4.2379699 &   0.083192 &   4.4638  & 81.173 &  2$f_0$  \\
  6.3567948 &   0.025942 &   2.5909  & 69.957 &  3$f_0$  \\
  8.4750872 &   0.003859 &   0.9586  & 25.783 &  4$f_0$  \\
 10.5955076 &   0.004677 &   2.0603  & 26.972 &  5$f_0$  \\
 12.7142172 &   0.005179 &   0.1356  & 32.102 &  6$f_0$  \\
 14.8329821 &   0.003394 &   4.4305  & 24.368 &  7$f_0$  \\
 16.9521809 &   0.002134 &   2.1695  & 16.597 &  8$f_0$  \\
 19.0711250 &   0.000924 &   0.0034  &  9.907 &  9$f_0$    \\
 21.1895828 &   0.000362 &   2.6349  &  5.089 &  10$f_0$  \\
 23.3090458 &   0.000650 &   0.1582  &  8.902 &  11$f_0$  \\
 25.4280815 &   0.000657 &   4.1858  &  9.371 &  12$f_0$  \\
 27.5470715 &   0.000427 &   2.1154  &  5.925 &  13$f_0$  \\
\\
\multicolumn{5}{l}{Blazhko modulation}\\
\\
  0.0557860 &   0.002250 &   3.5242  & 11.027 &   $f_m$  \\
  0.1130970 &   0.005032 &   2.0600  & 21.865 &   2$f_m$  \\

\\
\multicolumn{5}{l}{Modulation triplet frequencies}\\
\\
  2.0706120 &   0.001003 &   4.3047  &  6.254 &  $f_0-f_m$  \\
  2.1719539 &   0.004580 &   5.3412  & 17.976 &  $f_0+f_m$  \\
  4.1836190 &   0.000864 &   4.5330  &  5.714 &  2$f_0-f_m$  \\
  4.2913609 &   0.004452 &   2.8623  & 22.069 &  2$f_0+f_m$  \\
  6.3063798 &   0.000914 &   1.7760  &  7.603 &  3$f_0-f_m$  \\
  6.4105959 &   0.003799 &   1.2182  & 19.970 &  3$f_0+f_m$  \\
  8.4165525 &   0.000603 &   2.4485  &  5.293 &  4$f_0-f_m$  \\
  8.5297117 &   0.002889 &   6.1076  & 18.665 &  4$f_0+f_m$  \\
 10.5445929 &   0.000403 &   4.8041  &  3.957 &  5$f_0-f_m$  \\
 10.6492252 &   0.001437 &   4.1638  & 10.704 &  5$f_0+f_m$  \\
 12.7694759 &   0.000618 &   2.3372  &  6.028 &  6$f_0+f_m$  \\
 17.0050011 &   0.000589 &   2.2322  &  7.826 &  8$f_0+f_m$  \\
 19.1245899 &   0.000586 &   0.3344  &  6.950 &  9$f_0+f_m$  \\
 21.2440968 &   0.000399 &   4.5191  &  5.034 & 10$f_0+f_m$  \\
\\
\multicolumn{5}{l}{Additional terms identified as independent modes}\\
\\

  3.1588659 &   0.002119 &   2.9665  &  9.251 &  $f_2$  \\
  3.6308789 &   0.002844 &   1.0416  & 17.450 &  $f_1$  \\
\\
\multicolumn{5}{l}{Linear combinations between additional terms and $f_0$,}\\ 
\multicolumn{5}{l}{with or without $f_m$} \\
\\
 
  0.5557140 &   0.000579 &   1.4162  &  3.590 & $2f_0-f_m-f_1$  \\
  0.6137300 &   0.000967 &   4.3577  &  5.458 & $2f_0-f_1$  \\
  0.9815810 &   0.000673 &   5.9134  &  4.446 & $f_2-f_0-f_m$  \\
  1.0310810 &   0.000907 &   4.4587  &  5.680 & $2f_0-f_m-f_2$  \\
  1.0892100 &   0.000625 &   0.2036  &  4.373 & $2f_0-f_2$  \\
  1.1364660 &   0.000529 &   0.0723  &  4.307 & $2f_0+f_m-f_2$  \\
  1.5060490 &   0.001430 &   3.0864  &  9.914 & $f_1-f_0$  \\
  2.7380049 &   0.000817 &   0.5867  &  4.017 & $3f_0-f_1$  \\
  3.0997059 &   0.001693 &   2.0588  &  7.816 & $f_2-f_m$ \\
  3.2565880 &   0.000691 &   5.0736  &  4.663 & $3f_0+f_m-f_2$  \\
  4.8528800 &   0.000808 &   4.6690  &  5.310 & $4f_0-f_1$  \\
  5.1331830 &   0.000699 &   1.0265  &  4.088 & $2f_1-f_0$  \\
  5.2227440 &   0.000559 &   0.6052  &  4.755 & $f_0-f_m+f_2$  \\
  5.2794271 &   0.000596 &   6.1065  &  4.413 & $f_0+f_2$  \\
  5.3320541 &   0.000404 &   0.1849  &  3.706 & $f_0+f_m+f_2$  \\
  5.3743072 &   0.000564 &   2.6932  &  5.144 & $4f_0+f_m-f_2$  \\
  5.6918921 &   0.000502 &   1.0524  &  3.988 & $f_0-f_m+f_1$  \\
  5.7496319 &   0.002907 &   5.7320  & 18.811 & $f_0+f_1$  \\
  6.9680071 &   0.000622 &   3.9101  &  4.517 & $5f_0-f_1$  \\
  7.2510352 &   0.000450 &   3.3737  &  3.567 &  $2f_1$  \\
  7.3422809 &   0.000408 &   6.1252  &  3.540 & $2f_0-f_m+f_2$  \\
  7.3839550 &   0.000509 &   2.9947  &  4.455 & $5f_0-f_m+f_2$  \\
  7.3946028 &   0.001058 &   5.1992  &  8.021 & $2f_0+f_2$  \\
  7.8696060 &   0.001498 &   3.8452  & 10.900 & $2f_0+f_1$  \\
  8.9037333 &   0.000488 &   1.7403  &  5.396 & $f_0+f_1+f_2$  \\
  8.9577160 &   0.000431 &   3.1197  &  4.556 & $f_0+f_m+f_1+f_2$  \\
  9.0911274 &   0.000523 &   1.7513  &  4.075 & $6f_0-f_1$  \\
  9.3717070 &   0.000840 &   2.8583  &  5.426 & $f_0+2f_1$  \\
  9.4592342 &   0.000536 &   4.5554  &  4.219 & $3f_0-f_m+f_2$  \\
  9.5042696 &   0.000520 &   1.3052  &  4.413 & $6f_0-f_m-f_2$  \\
  9.5122747 &   0.001021 &   3.4523  &  7.300 & $3f_0+f_2$  \\
  9.9870539 &   0.002025 &   2.1231  & 14.298 & $3f_0+f_1$  \\
 10.0394773 &   0.000660 &   4.7210  &  5.748 & $3f_0+f_m+f_1$  \\
 11.4905624 &   0.000896 &   0.9388  &  6.940 & $2f_0+2f_1$  \\
 11.5774755 &   0.000494 &   2.9245  &  4.427 & $4f_0-f_m+f_2$  \\
 11.6231632 &   0.000689 &   6.1547  &  6.080 & $7f_0-f_m-f_2$  \\  
 11.6310463 &   0.000901 &   1.6964  &  7.590 & $4f_0+f_2$  \\
 12.1040316 &   0.001536 &   0.2982  & 12.757 & $4f_0+f_1$  \\
 12.1594105 &   0.000564 &   3.3520  &  5.916 & $4f_0+f_m+f_1$  \\
 13.6098433 &   0.000621 &   5.4812  &  5.738 & $3f_0+2f_1$  \\
 13.7422590 &   0.000518 &   4.7605  &  5.055 & $8f_0-f_m-f_2$  \\
 13.7495470 &   0.000543 &   5.8919  &  5.436 & $5f_0+f_2$  \\
 14.2228832 &   0.001424 &   4.3996  & 12.419 & $5f_0+f_1$  \\
 14.2800894 &   0.000424 &   1.1744  &  5.263 & $5f_0+f_m+f1$  \\
 15.8649635 &   0.000590 &   3.1909  &  6.673 & $9f_0-f_m-f_2$  \\
 16.3416233 &   0.001096 &   2.4098  & 11.503 & $6f_0+f_1$  \\
 18.4694328 &   0.000364 &   2.2220  &  4.735 & $7f_0+f_1$  \\
 20.5785885 &   0.000401 &   5.4055  &  5.082 & $8f_0+f_1$  \\

%% file: 14941tb2.tex
  1  &   4237.1460  &  --0.3481  & $  -0.0011 $   &&152  &   4308.4114  &  --0.3176  & $   0.0039   $ \\
  2  &   4237.6211  &  --0.3223  & $   0.0020 $   &&153  &   4308.8767  &  --0.3193  & $  -0.0026   $ \\
  3  &   4238.0891  &  --0.3212  & $  -0.0019 $   &&154  &   4309.3486  &  --0.3275  & $  -0.0027   $ \\
  4  &   4238.5593  &  --0.3433  & $  -0.0036 $   &&155  &   4309.8227  &  --0.3256  & $  -0.0005   $ \\
  5  &   4239.0325  &  --0.3235  & $  -0.0023 $   &&156  &   4310.2949  &  --0.3154  & $  -0.0002   $ \\
  6  &   4239.5027  &  --0.3198  & $  -0.0040 $   &&157  &   4310.7649  &  --0.3407  & $  -0.0022   $ \\
  7  &   4239.9727  &  --0.3320  & $  -0.0060 $   &&158  &   4311.2380  &  --0.3426  & $  -0.0009   $ \\
  8  &   4240.4509  &  --0.3178  & $   0.0003 $   &&159  &   4311.7131  &  --0.3095  & $   0.0022   $ \\
  9  &   4240.9238  &  --0.2999  & $   0.0013 $   &&160  &   4312.1831  &  --0.3143  & $   0.0003   $ \\
 10  &   4241.3940  &  --0.3041  & $  -0.0004 $   &&161  &   4312.6543  &  --0.3347  & $  -0.0004   $ \\
 11  &   4241.8623  &  --0.3200  & $  -0.0040 $   &&162  &   4313.1284  &  --0.3271  & $   0.0018   $ \\
 12  &   4242.3445  &  --0.2937  & $   0.0062 $   &&163  &   4313.5996  &  --0.3015  & $   0.0010   $ \\
 13  &   4242.8106  &  --0.2937  & $   0.0004 $   &&164  &   4314.0696  &  --0.3206  & $  -0.0009   $ \\
 14  &   4243.2808  &  --0.2988  & $  -0.0014 $   &&165  &   4314.5408  &  --0.3260  & $  -0.0017   $ \\
 15  &   4243.7578  &  --0.2943  & $   0.0038 $   &&166  &   4315.0198  &  --0.3010  & $   0.0054   $ \\
 16  &   4244.2319  &  --0.2883  & $   0.0060 $   &&167  &   4315.4849  &  --0.3081  & $  -0.0014   $ \\
 17  &   4244.6968  &  --0.2926  & $  -0.0011 $   &&168  &   4315.9570  &  --0.3149  & $  -0.0012   $ \\
 18  &   4245.1719  &  --0.3093  & $   0.0021 $   &&169  &   4316.4329  &  --0.3228  & $   0.0027   $ \\
 19  &   4245.6462  &  --0.2897  & $   0.0045 $   &&170  &   4316.9050  &  --0.3029  & $   0.0030   $ \\
 20  &   4246.1162  &  --0.2878  & $   0.0026 $   &&171  &   4317.3694  &  --0.3174  & $  -0.0046   $ \\
 21  &   4246.5864  &  --0.3196  & $   0.0009 $   &&172  &   4317.8484  &  --0.3234  & $   0.0025   $ \\
 22  &   4247.0586  &  --0.3165  & $   0.0011 $   &&173  &   4318.3193  &  --0.3075  & $   0.0015   $ \\
 23  &   4247.5308  &  --0.2958  & $   0.0013 $   &&174  &   4318.7915  &  --0.2968  & $   0.0018   $ \\
 24  &   4248.0059  &  --0.3138  & $   0.0045 $   &&175  &   4319.2605  &  --0.3319  & $  -0.0012   $ \\
 25  &   4248.4749  &  --0.3327  & $   0.0016 $   &&176  &   4319.7356  &  --0.3056  & $   0.0020   $ \\
 26  &   4248.9490  &  --0.3107  & $   0.0038 $   &&177  &   4320.2078  &  --0.2931  & $   0.0023   $ \\
 27  &   4249.4204  &  --0.3115  & $   0.0033 $   &&178  &   4320.6777  &  --0.3057  & $   0.0003   $ \\
 28  &   4249.8914  &  --0.3254  & $   0.0023 $   &&179  &   4321.1499  &  --0.3289  & $   0.0006   $ \\
 29  &   4250.3613  &  --0.3306  & $   0.0004 $   &&180  &   4321.6282  &  --0.2922  & $   0.0069   $ \\
 30  &   4250.8367  &  --0.3198  & $   0.0038 $   &&181  &   4322.0920  &  --0.3153  & $  -0.0012   $ \\
 31  &   4251.3047  &  --0.3351  & $  -0.0001 $   &&182  &   4322.5652  &  --0.3069  & $   0.0001   $ \\
 32  &   4251.7788  &  --0.3297  & $   0.0021 $   &&183  &   4323.0403  &  --0.3153  & $   0.0032   $ \\
 33  &   4252.2522  &  --0.3181  & $   0.0036 $   &&184  &   4323.5112  &  --0.3005  & $   0.0023   $ \\
 34  &   4252.7212  &  --0.3187  & $   0.0006 $   &&185  &   4323.9815  &  --0.3187  & $   0.0006   $ \\
 35  &   4253.1924  &  --0.3417  & $  -0.0001 $   &&186  &   4324.4536  &  --0.3198  & $   0.0008   $ \\
 36  &   4253.6655  &  --0.3186  & $   0.0011 $   &&187  &   4324.9316  &  --0.3000  & $   0.0069   $ \\
 37  &   4254.1375  &  --0.3269  & $   0.0011 $   &&188  &   4325.3948  &  --0.3226  & $  -0.0019   $ \\
 38  &   4254.6047  &  --0.3417  & $  -0.0035 $   &&189  &   4325.8718  &  --0.3295  & $   0.0033   $ \\
 39  &   4255.0808  &  --0.3360  & $   0.0006 $   &&190  &   4326.3428  &  --0.3055  & $   0.0023   $ \\
 40  &   4255.5510  &  --0.3204  & $  -0.0011 $   &&191  &   4326.8140  &  --0.3040  & $   0.0015   $ \\
 41  &   4256.0230  &  --0.3243  & $  -0.0011 $   &&192  &   4327.2830  &  --0.3466  & $  -0.0014   $ \\
 42  &   4256.4919  &  --0.3385  & $  -0.0040 $   &&193  &   4327.7571  &  --0.3118  & $   0.0008   $ \\
 43  &   4256.9724  &  --0.3264  & $   0.0045 $   &&194  &   4328.2302  &  --0.3117  & $   0.0020   $ \\
 44  &   4257.4365  &  --0.3199  & $  -0.0033 $   &&195  &   4328.6973  &  --0.3270  & $  -0.0028   $ \\
 45  &   4257.9136  &  --0.3298  & $   0.0019 $   &&196  &   4329.1724  &  --0.3345  & $   0.0003   $ \\
 46  &   4258.3826  &  --0.3174  & $  -0.0011 $   &&197  &   4329.6475  &  --0.3096  & $   0.0035   $ \\
 47  &   4258.8530  &  --0.3175  & $  -0.0025 $   &&198  &   4330.1125  &  --0.3209  & $  -0.0033   $ \\
 48  &   4259.3240  &  --0.3164  & $  -0.0035 $   &&199  &   4330.5867  &  --0.3393  & $  -0.0011   $ \\
 49  &   4259.8010  &  --0.3153  & $   0.0016 $   &&200  &   4331.0588  &  --0.3096  & $  -0.0009   $ \\
 50  &   4260.2703  &  --0.3012  & $  -0.0011 $   &&201  &   4331.5337  &  --0.2995  & $   0.0020   $ \\
 51  &   4260.7444  &  --0.3005  & $   0.0011 $   &&202  &   4332.0019  &  --0.3159  & $  -0.0016   $ \\
 52  &   4261.2146  &  --0.3120  & $  -0.0006 $   &&203  &   4332.4749  &  --0.3266  & $  -0.0006   $ \\
 53  &   4261.6899  &  --0.2990  & $   0.0028 $   &&204  &   4332.9490  &  --0.2930  & $   0.0016   $ \\
 54  &   4262.1587  &  --0.2893  & $  -0.0003 $   &&205  &   4333.4202  &  --0.3034  & $   0.0008   $ \\
 55  &   4262.6318  &  --0.3043  & $   0.0009 $   &&206  &   4333.8892  &  --0.3261  & $  -0.0021   $ \\
 56  &   4263.1050  &  --0.3056  & $   0.0021 $   &&207  &   4334.3662  &  --0.3104  & $   0.0030   $ \\
 57  &   4263.5791  &  --0.3029  & $   0.0043 $   &&208  &   4334.8352  &  --0.2882  & $   0.0001   $ \\
 58  &   4264.0471  &  --0.3028  & $   0.0004 $   &&209  &   4335.3103  &  --0.3210  & $   0.0033   $ \\
 59  &   4264.5215  &  --0.3199  & $   0.0028 $   &&210  &   4335.7776  &  --0.3119  & $  -0.0014   $ \\
 60  &   4264.9954  &  --0.3087  & $   0.0048 $   &&211  &   4336.2585  &  --0.2975  & $   0.0077   $ \\
 61  &   4265.4658  &  --0.3106  & $   0.0033 $   &&212  &   4336.7236  &  --0.3023  & $   0.0008   $ \\
 62  &   4265.9358  &  --0.3246  & $   0.0014 $   &&213  &   4337.1985  &  --0.2960  & $   0.0038   $ \\
 63  &   4266.4131  &  --0.3258  & $   0.0068 $   &&214  &   4337.6697  &  --0.3065  & $   0.0030   $ \\
 64  &   4266.8801  &  --0.3175  & $   0.0019 $   &&215  &   4338.1409  &  --0.3007  & $   0.0023   $ \\
 65  &   4267.3501  &  --0.3299  & $  -0.0001 $   &&216  &   4338.6108  &  --0.3005  & $   0.0004   $ \\
 66  &   4267.8262  &  --0.3299  & $   0.0041 $   &&217  &   4339.0862  &  --0.2863  & $   0.0038   $ \\
 67  &   4268.2986  &  --0.3249  & $   0.0046 $   &&218  &   4339.5559  &  --0.3066  & $   0.0016   $ \\
 68  &   4268.7656  &  --0.3077  & $  -0.0003 $   &&219  &   4340.0310  &  --0.2875  & $   0.0048   $ \\
 69  &   4269.2427  &  --0.3302  & $   0.0048 $   &&220  &   4340.5002  &  --0.3037  & $   0.0021   $ \\
 70  &   4269.7126  &  --0.3390  & $   0.0029 $   &&221  &   4340.9712  &  --0.2917  & $   0.0011   $ \\
 71  &   4270.1858  &  --0.3101  & $   0.0041 $   &&222  &   4341.4473  &  --0.2889  & $   0.0053   $ \\
 72  &   4270.6531  &  --0.3371  & $  -0.0005 $   &&223  &   4341.9175  &  --0.3070  & $   0.0035   $ \\
 73  &   4271.1282  &  --0.3407  & $   0.0026 $   &&224  &   4342.3852  &  --0.3147  & $  -0.0006   $ \\
 74  &   4271.6013  &  --0.3368  & $   0.0038 $   &&225  &   4342.8628  &  --0.2995  & $   0.0050   $ \\
 75  &   4272.0684  &  --0.3376  & $  -0.0010 $   &&226  &   4343.3315  &  --0.3275  & $   0.0018   $ \\
 76  &   4272.5435  &  --0.3503  & $   0.0021 $   &&227  &   4343.8008  &  --0.3282  & $  -0.0009   $ \\
 77  &   4273.0137  &  --0.3427  & $   0.0004 $   &&228  &   4344.2766  &  --0.3247  & $   0.0031   $ \\
 78  &   4273.4856  &  --0.3354  & $   0.0004 $   &&229  &   4344.7488  &  --0.3029  & $   0.0033   $ \\
 79  &   4273.9539  &  --0.3491  & $  -0.0032 $   &&230  &   4345.2161  &  --0.3364  & $  -0.0013   $ \\
 80  &   4274.4290  &  --0.3353  & $   0.0000 $   &&231  &   4345.6919  &  --0.3239  & $   0.0026   $ \\
 81  &   4274.9001  &  --0.3355  & $  -0.0008 $   &&232  &   4346.1650  &  --0.3162  & $   0.0038   $ \\
 82  &   4275.3665  &  --0.3333  & $  -0.0064 $   &&233  &   4346.6331  &  --0.3149  & $  -0.0001   $ \\
 83  &   4275.8435  &  --0.3415  & $  -0.0013 $   &&234  &   4347.1062  &  --0.3253  & $   0.0011   $ \\
 84  &   4276.3203  &  --0.3182  & $   0.0036 $   &&235  &   4347.5784  &  --0.3026  & $   0.0014   $ \\
 85  &   4276.7856  &  --0.3190  & $  -0.0030 $   &&236  &   4348.0503  &  --0.3075  & $   0.0014   $ \\
 86  &   4277.2598  &  --0.3150  & $  -0.0008 $   &&237  &   4348.5215  &  --0.3087  & $   0.0006   $ \\
 87  &   4277.7339  &  --0.3124  & $   0.0014 $   &&238  &   4348.9944  &  --0.3202  & $   0.0016   $ \\
 88  &   4278.2031  &  --0.3145  & $  -0.0013 $   &&239  &   4349.4675  &  --0.2917  & $   0.0028   $ \\
 89  &   4278.6721  &  --0.3176  & $  -0.0042 $   &&240  &   4349.9358  &  --0.3186  & $  -0.0008   $ \\
 90  &   4279.1523  &  --0.3063  & $   0.0041 $   &&241  &   4350.4087  &  --0.2909  & $   0.0001   $ \\
 91  &   4279.6213  &  --0.3028  & $   0.0012 $   &&242  &   4350.8818  &  --0.3119  & $   0.0014   $ \\
 92  &   4280.0874  &  --0.3091  & $  -0.0047 $   &&243  &   4351.3518  &  --0.2884  & $  -0.0006   $ \\
 93  &   4280.5644  &  --0.3216  & $   0.0004 $   &&244  &   4351.8259  &  --0.3056  & $   0.0016   $ \\
 94  &   4281.0405  &  --0.3044  & $   0.0046 $   &&245  &   4352.2949  &  --0.3014  & $  -0.0013   $ \\
 95  &   4281.5068  &  --0.3080  & $  -0.0010 $   &&246  &   4352.7739  &  --0.3039  & $   0.0058   $ \\
 96  &   4281.9761  &  --0.3297  & $  -0.0037 $   &&247  &   4353.2380  &  --0.2996  & $  -0.0020   $ \\
 97  &   4282.4570  &  --0.3057  & $   0.0053 $   &&248  &   4353.7141  &  --0.3066  & $   0.0021   $ \\
 98  &   4282.9250  &  --0.3211  & $   0.0014 $   &&249  &   4354.1863  &  --0.2979  & $   0.0024   $ \\
 99  &   4283.3914  &  --0.3128  & $  -0.0042 $   &&250  &   4354.6611  &  --0.2951  & $   0.0053   $ \\
100  &   4283.8684  &  --0.3320  & $   0.0010 $   &&251  &   4355.1282  &  --0.2945  & $   0.0004   $ \\
101  &   4284.3433  &  --0.3093  & $   0.0039 $   &&252  &   4355.6055  &  --0.2942  & $   0.0058   $ \\
102  &   4284.8106  &  --0.3158  & $  -0.0008 $   &&253  &   4356.0745  &  --0.2756  & $   0.0028   $ \\
103  &   4285.2837  &  --0.3251  & $   0.0005 $   &&254  &   4356.5464  &  --0.2858  & $   0.0028   $ \\
104  &   4285.7590  &  --0.3222  & $   0.0039 $   &&255  &   4357.0156  &  --0.2945  & $   0.0002   $ \\
105  &   4286.2270  &  --0.3157  & $   0.0000 $   &&256  &   4357.4951  &  --0.2857  & $   0.0077   $ \\
106  &   4286.6951  &  --0.3195  & $  -0.0039 $   &&257  &   4357.9607  &  --0.2790  & $   0.0014   $ \\
107  &   4287.1699  &  --0.3237  & $  -0.0010 $   &&258  &   4358.4358  &  --0.2837  & $   0.0046   $ \\
108  &   4287.6453  &  --0.3123  & $   0.0024 $   &&259  &   4358.9048  &  --0.2916  & $   0.0016   $ \\
109  &   4288.1155  &  --0.3056  & $   0.0007 $   &&260  &   4359.3784  &  --0.3010  & $   0.0033   $ \\
110  &   4288.5845  &  --0.3387  & $  -0.0022 $   &&261  &   4359.8520  &  --0.2877  & $   0.0051   $ \\
111  &   4289.0606  &  --0.3156  & $   0.0019 $   &&262  &   4360.3210  &  --0.3030  & $   0.0021   $ \\
112  &   4289.5298  &  --0.3262  & $  -0.0007 $   &&263  &   4360.7952  &  --0.3060  & $   0.0043   $ \\
113  &   4290.0010  &  --0.3242  & $  -0.0015 $   &&264  &   4361.2688  &  --0.3061  & $   0.0060   $ \\
114  &   4290.4749  &  --0.3425  & $   0.0005 $   &&265  &   4361.7363  &  --0.3104  & $   0.0016   $ \\
115  &   4290.9470  &  --0.3333  & $   0.0007 $   &&266  &   4362.2102  &  --0.3195  & $   0.0036   $ \\
116  &   4291.4163  &  --0.3356  & $  -0.0020 $   &&267  &   4362.6824  &  --0.3210  & $   0.0038   $ \\
117  &   4291.8882  &  --0.3453  & $  -0.0020 $   &&268  &   4363.1555  &  --0.3195  & $   0.0051   $ \\
118  &   4292.3633  &  --0.3340  & $   0.0012 $   &&269  &   4363.6216  &  --0.3224  & $  -0.0008   $ \\
119  &   4292.8335  &  --0.3353  & $  -0.0005 $   &&270  &   4364.0984  &  --0.3280  & $   0.0041   $ \\
120  &   4293.3005  &  --0.3438  & $  -0.0054 $   &&271  &   4364.5676  &  --0.3353  & $   0.0014   $ \\
121  &   4293.7817  &  --0.3115  & $   0.0039 $   &&272  &   4365.0418  &  --0.3224  & $   0.0036   $ \\
122  &   4294.2498  &  --0.3288  & $   0.0000 $   &&273  &   4365.5088  &  --0.3331  & $  -0.0013   $ \\
123  &   4294.7207  &  --0.3229  & $  -0.0010 $   &&274  &   4365.9849  &  --0.3106  & $   0.0029   $ \\
124  &   4295.1931  &  --0.3282  & $  -0.0005 $   &&275  &   4366.4539  &  --0.3248  & $  -0.0001   $ \\
125  &   4295.6692  &  --0.3057  & $   0.0037 $   &&276  &   4366.9280  &  --0.3071  & $   0.0021   $ \\
126  &   4296.1331  &  --0.3142  & $  -0.0044 $   &&277  &   4367.3980  &  --0.3096  & $   0.0002   $ \\
127  &   4296.6074  &  --0.3194  & $  -0.0020 $   &&278  &   4367.8711  &  --0.3047  & $   0.0014   $ \\
128  &   4297.0815  &  --0.3070  & $   0.0002 $   &&279  &   4368.3411  &  --0.3370  & $  -0.0005   $ \\
129  &   4297.5554  &  --0.3018  & $   0.0022 $   &&280  &   4368.8142  &  --0.3108  & $   0.0007   $ \\
130  &   4298.0207  &  --0.3223  & $  -0.0044 $   &&281  &   4369.2864  &  --0.3170  & $   0.0009   $ \\
131  &   4298.4968  &  --0.3127  & $  -0.0002 $   &&282  &   4369.7617  &  --0.3135  & $   0.0043   $ \\
132  &   4298.9736  &  --0.2974  & $   0.0046 $   &&283  &   4370.2273  &  --0.3188  & $  -0.0020   $ \\
133  &   4299.4380  &  --0.3152  & $  -0.0029 $   &&284  &   4370.7043  &  --0.2918  & $   0.0031   $ \\
134  &   4299.9111  &  --0.3286  & $  -0.0017 $   &&285  &   4371.1753  &  --0.3181  & $   0.0021   $ \\
135  &   4300.3899  &  --0.2974  & $   0.0051 $   &&286  &   4371.6414  &  --0.2981  & $  -0.0037   $ \\
136  &   4300.8528  &  --0.2922  & $  -0.0039 $   &&287  &   4372.1194  &  --0.3154  & $   0.0024   $ \\
137  &   4301.3232  &  --0.3135  & $  -0.0054 $   &&288  &   4372.5916  &  --0.2918  & $   0.0026   $ \\
138  &   4301.8042  &  --0.3152  & $   0.0037 $   &&289  &   4373.0647  &  --0.2853  & $   0.0039   $ \\
139  &   4302.2751  &  --0.3037  & $   0.0027 $   &&290  &   4373.5308  &  --0.3080  & $  -0.0020   $ \\
140  &   4302.7415  &  --0.3143  & $  -0.0029 $   &&291  &   4374.0107  &  --0.2919  & $   0.0061   $ \\
141  &   4303.2144  &  --0.3106  & $  -0.0019 $   &&292  &   4374.4810  &  --0.2765  & $   0.0044   $ \\
142  &   4303.6917  &  --0.2957  & $   0.0034 $   &&293  &   4374.9490  &  --0.2911  & $   0.0004   $ \\
143  &   4304.1616  &  --0.3063  & $   0.0015 $   &&294  &   4375.4228  &  --0.2970  & $   0.0024   $ \\
144  &   4304.6328  &  --0.3113  & $   0.0008 $   &&295  &   4375.8992  &  --0.2899  & $   0.0068   $ \\
145  &   4305.1047  &  --0.3109  & $   0.0008 $   &&296  &   4376.3660  &  --0.2824  & $   0.0017   $ \\
146  &   4305.5779  &  --0.2979  & $   0.0020 $   &&297  &   4376.8420  &  --0.3036  & $   0.0058   $ \\
147  &   4306.0440  &  --0.3199  & $  -0.0039 $   &&298  &   4377.3123  &  --0.2884  & $   0.0041   $ \\
148  &   4306.5190  &  --0.3092  & $  -0.0007 $   &&299  &   4377.7832  &  --0.2861  & $   0.0031   $ \\
149  &   4306.9951  &  --0.3014  & $   0.0035 $   &&300  &   4378.2534  &  --0.2970  & $   0.0014   $ \\
150  &   4307.4644  &  --0.3171  & $   0.0008 $   &&301  &   4378.7258  &  --0.3097  & $   0.0019   $ \\
151  &   4307.9333  &  --0.3326  & $  -0.0022 $  &&&& \\